\documentclass[amsmath,twocolumn,prb,aps]{revtex4-1}
\usepackage{amsthm,amsfonts,graphicx,verbatim, color}
\usepackage{bm}% bold math
\usepackage[utf8]{inputenc}
\usepackage{amsmath, amssymb}
\usepackage[caption=false]{subfig}
\usepackage{booktabs}
\usepackage{tikz}
\newcommand{\mb}[1]{\ensuremath{\mathbf{#1}} }
\newcommand{\mc}[1]{\ensuremath{\mathcal{#1}} }

\newcommand{\bra}[1]{\ensuremath{\langle #1 |}}
\newcommand{\ket}[1]{\ensuremath{| #1 \rangle}}
\newcommand{\LL}{\mathcal L}

\newcommand*{\citen}[1]{%
  \begingroup
    \romannumeral-`\x % remove space at the beginning of \setcitestyle
    \setcitestyle{numbers}%
    \cite{#1}%
  \endgroup   
}

\begin{document}
\def\bbm[#1]{\mbox{\boldmath$#1$}}

\title{Integer quantum Hall transition in a {\it fraction} of a Landau level} 

\author{Matteo Ippoliti, Scott D. Geraedts, and R. N. Bhatt}

\affiliation{Departments of Electrical Engineering and Physics, Princeton University, Princeton NJ 08544, USA}

\begin{abstract}
We investigate the quantum Hall problem in the lowest Landau level in two dimensions, in the presence of an arbitrary number of $\delta$-function potentials arranged in different geometric configurations. 
When the number of delta functions $N_\delta$ is smaller than the number of flux quanta through the system ($N_\phi$), there is a manifold of $(N_\phi-N_\delta)$ degenerate states at the original Landau level energy. 
We prove that the total Chern number of this set of states is +1 regardless of the number or position of the $\delta$ functions. 
Furthermore, we find numerically that, upon the addition of disorder, this subspace includes a quantum Hall transition which is (in a well-defined sense) {\it quantitatively} the same as that for the lowest Landau level without $\delta$-function impurities, but with a reduced number $N_\phi' \equiv N_\phi-N_\delta$ of magnetic flux quanta. 
We discuss the implications of these results for studies of the integer plateau transitions, as well as for the many-body problem in the presence of electron-electron interactions. 
\end{abstract}

\maketitle

\section{Introduction \label{sec:intro}}

A crucial part in the understanding of the integer quantum Hall effect\citep{vonklitzing_1980} was the realization that the Hall conductance is robust to the presence of impurities.
Before this robustness was proved for general disorder\cite{laughlin_1981} and connected to topology\cite{TKNN}, Prange\citep{prange_1981} showed that
if a single impurity represented by a $\delta$-function potential is placed in a quantum Hall system, the Hall conductance does not change.
Even though one electron's cyclotron orbit becomes bound to the impurity 
(as has recently been shown\cite{feldman_2016} in spectacular STM images of surface states of Bi in a strong magnetic field), 
the remaining electrons exactly compensate the loss of conductance due to the presence of the bound state.

In this paper, we consider a generalized case in which an arbitrary number $N_\delta$ of impurities represented by $\delta$-function potentials (which we call ``$\delta$-impurities'' in the following) is placed in a 2D electron gas pierced by $N_\phi$ magnetic flux quanta.
We find that the same conclusion as Prange's single-impurity case applies: 
there is a set of states which are not bound by the $\delta$-impurities, as was also predicted by Brezin {\it et al.} \citep{brezin_1984}, 
and these states carry the quantized Hall current corresponding to the entire Landau level.
In other words, we show that the subspace of electron states which avoid all $\delta$-impurities is characterized by a Chern number \cite{hatsugai_1993} $C = +1$.
We thus have a subspace of \textit{degenerate} electron states with $C=1$,
analogous to a Landau level of reduced dimension $N_{\phi}' \equiv N_\phi-N_\delta$. 
As such, this ``fraction'' of a Landau level should exhibit an integer quantum Hall transition.
Projecting onto this lower dimensional space, which can be arbitrarily smaller than the original Landau level subspace, would seem to offer the possibility of studying quantum Hall transitions for system sizes that are much larger than those possible for the full Landau level problem. Based on that hope, we have carried out a study of Hall as well Thouless (longitudinal) conductance for varying degrees of ``dilution'' of the Landau level Hilbert space and various geometric distributions of $\delta$-impurities.

With a lattice of $\delta$-impurities with identical or periodically varying strength, 
in the regime $N_\delta < N_\phi$, one has, in addition to the flat $C=1$ band, one or more dispersive bands with varying Chern character, depending on the nature of the lattice.
This allows one to create Chern insulator models of different kinds, which have been of increasing interest in tight binding models; 
here they arise out of a single Landau level.
Thus this model may also offer a rich variety of phases upon addition of electron-electron interactions, such as fractional Chern insulators\cite{spanton_2017}, as well as the possibility of many-body localization, which appears to not be possible in a single Landau level subject to a random (e.g. white-noise) potential characterized by a single disorder strength \citep{nandkishore_2014, geraedts_2016}.

Our paper is structured as follows.
In Sec.~\ref{sec:pot} we present general facts about $\delta$-function potentials in the lowest Landau level, including the existence of $(N_\phi-N_\delta)$ degenerate states at the original Landau level energy.
In Sec.~\ref{sec:chern} we discuss the total Hall conductance of said subspace and show that its total Chern number is $+1$.
Sec.~\ref{sec:delta_fcn} discusses $\delta$-function potentials with lattice symmetry and the structure of the subbands they give rise to.
In Sec.~\ref{sec:numerics} we present numerical calculations of the Hall and Thouless conductance in the presence of disorder, both with and without $\delta$-impurities, and show that the quantum Hall transition seems to only depend on the number of states left at the Landau level energy, while being completely unaffected by the states localized by the $\delta$-impurities.
Finally, we summarize our results and discuss avenues for future research in Sec.~\ref{sec:conclusion}.

\section{Delta-function potentials in the lowest Landau level \label{sec:pot}}

We consider a two-dimensional electron system in a strong magnetic field, so that the cyclotron gap in infinite for practical purposed.
We add a set of point impurities with $\delta$-function potentials, so that the Hamiltonian within the lowest Landau level is
$H = \frac{1}{2} \hbar \omega_c + V(\mb r)$, with
\begin{equation}
V(\mb r) =  \sum_{i=1}^{N_\delta} \lambda_i \delta(\mb r - \mb r_i)\;.
\end{equation}
Here $\{\mb r_i:\ i=1,\dots N_\delta \}$ are the positions of the impurities and $\{\lambda_i\}$ denote the strength of each impurity.
In the following we discard the constant energy shift $\frac{1}{2} \hbar \omega_c$ and refer to the Landau level energy as ``zero energy'', or $E=0$, for simplicity.

Let $\{ | \psi_n \rangle:\ n=1,\dots N_\phi \}$ be an orthonormal basis of states in the lowest Landau level. 
The matrix element of $V$ between any two such basis states is
\begin{equation}
V_{mn} = \langle \psi_m | V(\mb r) | \psi_n \rangle
= \sum_i \lambda_i \psi_m^*(\mb r_i) \psi_n(\mb r_i) 
\label{eq:Vmn}
\end{equation}
By defining the $N_\delta \times N_\phi$ matrix $v_{in} \equiv \psi_n(\mb r_i)$ and the $N_\delta \times N_\delta$ diagonal matrix $\Lambda_{ij} \equiv \lambda_i \delta_{ij}$,
\eqref{eq:Vmn} can be rewritten in matrix notation as
\begin{equation}
V = v^\dagger \Lambda v\;.
\label{eq:v_lambda_v}
\end{equation}
Thus the kernel of $V$ contains the kernel of $v$.
The latter consists of states that have vanishing amplitude on all impurities.
Indeed, consider one vector $\alpha$ in the kernel of $v$: this vector, viewed as a list of coefficients in the $\{ \ket{\psi_n}\}$ basis, defines a wavefunction $\psi(\mb r)$ such that
\begin{equation}
\psi(\mb r_i) = \sum_n \alpha_n \psi_n(\mb r_i) = \sum_n v_{in} \alpha_n= 0 \quad \forall \, i \;.
\end{equation}
There are $N_\phi - N_\delta$ independent states with this property (provided $N_\phi \geq N_\delta$). 
Generically these are the only zero-energy states present, i.e., for random values of the positions $\{ \mb r_i \}$ and strengths $\{ \lambda_i\}$ of the $\delta$-impurities, all other eigenvalues are non-zero with probability 1.
An important special case which we consider later in the paper is that of $\delta$-impurity {\it lattices}; 
in that case, it is possible to have zero-energy states even when $N_\delta > N_\phi$, but only if the potential is not of a definite sign. 
If $V$ is, for example, positive-definite (i.e. if all $\delta$-impurities are repulsive),
then an eigenstate $\psi_n$ that does {\it not} vanish on some impurity $i$ is such that
\begin{equation}
E_n = \bra{\psi_n} V \ket{\psi_n} \geq \lambda_i |\psi_n(\mb r_i)|^2 > 0\;.
\end{equation}
Thus if the $\delta$-function potential is positive-definite, it has $(N_\phi - N_\delta)$ zero-energy states which have vanishing amplitude on all impurities.
The remaining $N_\delta$ states have $E > 0$ and have non-zero amplitude on at least one impurity.

%%% introduce notation
Throughout the rest of the paper, we use the notation $\LL(N_\phi, N_\delta)$ to denote the kernel of a positive-definite potential with $N_\delta$ $\delta$-functions in the lowest Landau level of a system with $N_\phi$ flux quanta.
In particular, $\LL(N_\phi, 0)$ denotes the whole Hilbert space of a lowest Landau level with no impurities.
Based on the previous discussion, we have $\text{dim}\, \LL(N_\phi,N_\delta) = N_\phi - N_\delta$.

\section{Hall conductance of lowest Landau level with $\delta$-impurities \label{sec:chern}}

In this Section we compute the Chern number \cite{hatsugai_1993} of the completely filled $\LL(N_\phi, N_\delta)$ subspace and show that it always equals +1, regardless of the strength and position of the $\delta$-impurities, provided $N_\phi < N_\delta$.
For simplicity we consider a rectangular torus with sides $L_x$, $L_y$, but this assumption is not essential.

Consider the fully-occupied lowest Landau level $\LL(N_\phi,0)$. 
The corresponding many-electron wavefunction is obtained as a Slater determinant from single-electron wavefunctions and can be written in the Landau gauge as
\citep{haldane_1985}
\begin{equation}
\Psi(\{z_i\}) = e^{-\frac{1}{2} \sum_i y_i^2} F_{cm} (Z) \prod_{i<j} f(z_i-z_j) \;,
\end{equation}
where $f$ and $F_{cm}$ are holomorphic functions of the complex argument $z = x+iy$, $f$ is odd and $Z = \sum_i z_i$ is the ``center-of-mass'' coordinate. 
Quasiperiodicity can be used to constrain $F$ and $f$, and eventually this wavefunction can be used to prove the quantization of the Hall conductance for the $\nu=1$ integer quantum Hall effect.

In a similar way, we introduce the wavefunction for the completely filled $\LL(N_\phi, N_\delta)$ subspace.
This contains $N_\phi-N_\delta$ electrons which are constrained to avoid the impurity sites, $\{ \eta_i \}$.
We require the $\eta_i$'s to be non-degenerate but make no other assumptions about their spatial distribution.
The most general wavefunction for $N_{\phi}' \equiv N_\phi-N_\delta$ electrons in the lowest Landau level which vanishes on all impurities is given by
\begin{align}
\Psi_0 (\{z_i\}) 
& = \exp\left(-\frac{1}{2} \sum_{i=1}^{N_{\phi}'}  y_i^2 \right) F_{cm} (Z) \prod_{i<j=1}^{N_{\phi}'} f(z_i-z_j) \nonumber \\
& \qquad \times \prod_{i=1}^{N_{\phi}'} \prod_{j=1}^{N_\delta} f(z_i - \eta_j) \;.
\label{eq:Psi_0}
\end{align}
Following Ref.~\citen{haldane_1985}, we have that the most general form for $f$ is expressed in terms of Euler's Theta function 
$$ \vartheta_1(z|\tau) \equiv -\sum_{n \in \mathbb Z} e^{i\pi (n+1/2) + i\pi \tau (n+1/2)^2 + 2\pi i (n+1/2)z} $$
as
$f(z) = \vartheta_1 (z/L_x | iL_y/L_x)$.
Similarly, for the center-of-mass wavefunction we have
\begin{equation}
F_{cm} (Z) = e^{iKZ} \vartheta_1 \left( \frac{Z-Z_0}{L_x} \bigg| i\frac{L_y}{L_x} \right)
\end{equation}
so that the only remaining degrees of freedom in the ansatz are the quasi-momentum $K$ and the center-of-mass node $Z_0$.
By imposing generalized periodic boundary conditions with twist angles $\theta_x, \theta_y$, these are found to be
\begin{equation}
K = \frac{\theta_x}{L_x} - \frac{\pi}{L_x} (2b+N_\phi)  \label{eq:k}
\end{equation}
and
\begin{equation}
Z_0 = \frac{L_x}{2} \left( N_\phi + \frac{\theta_y}{\pi} + 2a \right) 
+ i N_\phi K - \sum_i \eta_i  \;,
\label{eq:z0} 
\end{equation}
where $a,b$ are integers chosen so that $Z_0$ is in the torus unit cell $-\frac{L_x}{2} \leq x < \frac{L_x}{2}$, $-\frac{L_y}{2} \leq y < \frac{L_y}{2}$.

It can be seen from Eq.~\eqref{eq:z0} that the node of the center-of-mass wavefunction, $Z_0$, can be moved to an arbitrary position on the torus by suitably adjusting the boundary twist angles $\theta_{x}$, $\theta_y$.
This is known to be a signature of non-zero Chern number\citep{bhatt_1988}.
However, we also prove that the $C = +1$ by a direct computation of the boundary integral
\begin{equation}
C = \frac{1}{2\pi i} \oint d\theta_i \langle \Psi_0 (\vec \theta) | \frac{\partial}{\partial \theta_i} | \Psi_0(\vec \theta) \rangle
\end{equation}
in Appendix~\ref{app:integral}.
This result holds for any spatial distribution of any number of $\delta$-impurities.
A much simpler proof for lattices of $\delta$-impurities is provided in the next Section.

\section{$\delta$-function lattice potentials}
\label{sec:delta_fcn}

\begin{figure}
\includegraphics[width = \columnwidth]{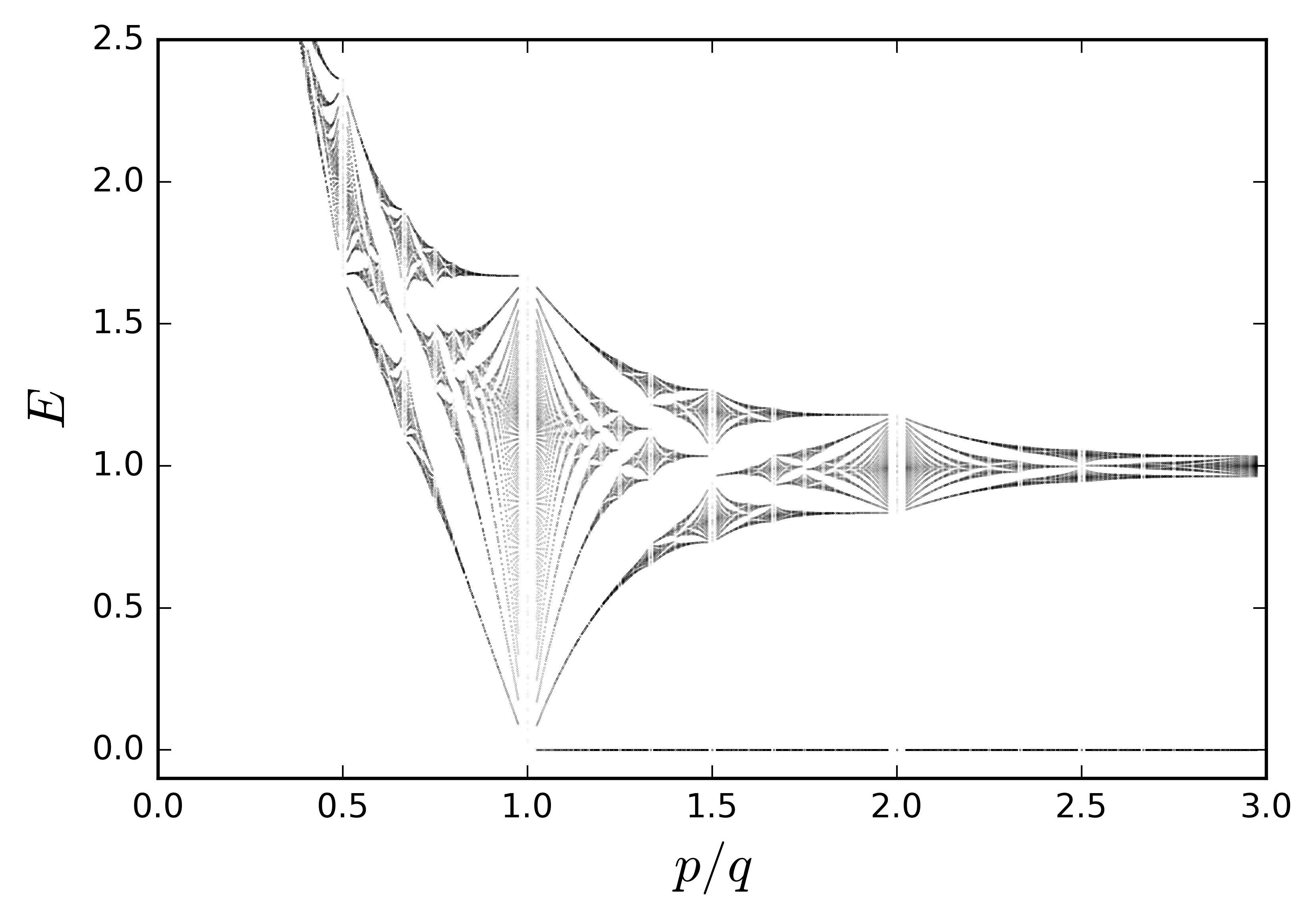}
\caption{
\label{fig:butterfly}
Spectrum of the $\delta$-function square lattice potential as a function of magnetic field.
The ratio $p/q$ equals the magnetic flux per unit cell, in units of the magnetic flux quantum.
It is also equal to $(2\pi \ell_B^2 n_\delta)^{-1}$, $n_\delta$ being the spatial density of $\delta$-impurities.
At small $p/q$, the electrons cannot resolve the individual $\delta$-impurities and see a nearly uniform potential of strength $n_\delta \sim q/p$, explaining the divergence of all energy levels.
At $p/q>1$, the zero-energy flat subband is clearly visible.
At large $p/q$, the $\delta$-impurities are so far apart that hopping is exponentially suppressed, thus the bandwitdth of all $E>0$ bands decays exponentially with $1/\sqrt{n_\delta}$. }
\end{figure}

In this Section we discuss $\delta$-function potentials with discrete translational symmetry, i.e. when the $\delta$-impurities are arranged on a lattice on the torus.
We assume a Bravais lattice generated by vectors $\mathbf a_1$, $\mathbf a_2$ (though it would be easy to generalize this to a lattice with a basis).
We further assume the torus has sides $\mathbf L_1 = N_1 \mathbf a_1$, $\mathbf L_2 = N_2 \mathbf a_2$, so that there is a total of $N_\delta = N_1 N_2$ $\delta$-impurities.
The potential is then given in real space by
\begin{equation}
V_\delta (\mathbf r) 
= \lambda \sum_{n_1 = 1}^{N_1} \sum_{n_2 = 1}^{N_2} 
 \delta(\mathbf r - n_1 \mathbf a_1 - n_2 \mathbf a_2 ) \;.
\label{eq:V_of_r}
\end{equation}
Furthermore, we assume that the magnetic flux through each unit cell is equal to $p/q$ quanta of magnetic flux, where $p$ and $q$ are co-prime integers with $p>q$. In other words, we require $\mathbf a_1 \times \mathbf a_2 = 2\pi \ell_B^2 p / q$.

The lattice symmetry allows us to pick a basis of eigenstates of $V_\delta$ which are also eigenstates of the magnetic translations \cite{zak_1965} $\hat \tau (q \mathbf a_1)$, $\hat \tau (\mathbf a_2)$ (the translations commute only if the enclosed area contains an integer number of flux quanta).
The eigenvalues of magnetic translations define a quasi-momentum $\mb k$, and the orbitals can be written in a quasi-Bloch form as
\begin{equation}
\psi_{\mb k,n}(\mb r) = e^{i\mb k \cdot \mb r} u_{\mb k,n}(\mb r) \;,
\end{equation}
where $n$ is a band index and the pseudo-Bloch wavefunction $u$ has the quasi-periodicity
\begin{equation}
\begin{aligned}
u_{\mb k,n}(\mb r+q \mathbf a_1) & = e^{-2\pi i (q\mathbf a_1) \times \mathbf r} u_{\mb k,n}(\mb r)\;, \\
u_{\mb k,n}(\mb r+ \mathbf a_2) & = u_{\mb k,n}(\mb r)\;.
\end{aligned}
\label{eq:u} 
\end{equation}
The matrix $V_\delta$ is then block-diagonalized into quasi-momentum sectors, with each block given by 
\begin{equation}
[V_\delta(\mb k)]_{nn'} = \bra{\psi_{\mb k,n}} V_\delta \ket{\psi_{\mb k, n'}}
\label{eq:vk_block}
\end{equation}
In Appendix~\ref{app:lattice} we discuss in detail the diagonalization of these potentials, which we then apply to our numerical calculations in Section~\ref{sec:lattice}.

We find that the Hamiltonian block \eqref{eq:vk_block} has the structure $V_\delta(\mb k) \propto v^\dagger(\mb k) v(\mb k)$, where $v(\mb k)$ is a rectangular $p \times q$ matrix.
Therefore, each Hamiltonian block is a $p\times p$ matrix of rank at most $q$, so $(p-q)$ eigenvalues are guaranteed to be zero.
Thus we find that there are $p$ subbands;
of these, $(p-q)$ are degenerate, flat, zero-energy subbands.
The spectrum of a square lattice of $\delta$-impurities as a function of the magnetic flux per unit cell $p/q$, showing the peculiar Hofstadter butterfly\cite{hofstadter_1976} fractal pattern, is shown in Fig.~\ref{fig:butterfly}.

The existence of this flat, zero-energy band is consistent with the general theory of periodic potentials in the lowest Landau level \citep{TKNN},
with previous studies of $\delta$-function lattices in the presence of a magnetic field\citep{gedik_1997,ishikawa_1995,ishikawa_1998,ishikawa_1999}, 
and with the discussion in the previous Sections about general $\delta$-function potentials.
Based on that discussion, we expect to have $(N_\phi - N_\delta)$ zero-energy states,
which is exactly the number of states contained in $(p-q)$ subbands (as each subband contains $N_\phi / p$ states).
Furthermore, based on the result of Section~\ref{sec:chern}, we know these subbands must carry a total Chern number of $+1$. 
This, too, is in agreement with known facts about periodic potentials in the lowest Landau level\cite{TKNN},
namely that the lowest $r$ subbands carry a total Chern number $C$ that solves the Diophantine equation
\begin{equation}
p C + q S = r
\label{eq:dioph}
\end{equation}
where $S$ is another integer.
Thus, the total Chern number of the lowest $p-q$ subbands, which in our case make up the flat zero-energy band, obeys $p(C-1) + q(S+1) = 0$.
The value $C = +1$ is always compatible with this constraint.
It is possible to prove that $C = +1$ by considering the number of states present in the zero-energy band, $(N_\phi - N_\delta)$; 
the Chern number can be computed as a derivative of this number with respect to $N_\phi$, with the lattice potential held constant\cite{streda_1982}:
\begin{equation}
C = \left. \frac{\partial (N_\phi - N_\delta)}{\partial N_\phi}\right|_{N_\delta} = +1 \;.
\end{equation}

These features of the subband structure have several interesting consequences.
First of all, we see from Eq.~\eqref{eq:dioph} that, by tuning $p$ and $q$, one can engineer subbands with large Chern numbers. 
%As an example, if $p/q = 11/10$, the allowed solutions of Eq.~\eqref{eq:dioph} are $C_j = 1+11k_j$, $k_j \in \mathbb Z$, and the total Landau level Chern number of $+1$ ensures that there is at least one $k_j \neq 0$, hence at least one subband with $|C_j| \geq 10$.
In the presence of interactions, these high-Chern-number subbands have the potential to host interesting strongly-correlated phases, such as the recently observed fractional Chern insulator states\cite{spanton_2017}.

Another interesting consequence is that, since the zero-energy subbands take up the entire Chern character of the Landau level, the remaining $q$ subbands taken together must have $C = 0$.
The simplest instance of this occurs for $q = 1$, when there is only one dispersive subband with bandwidth decreasing exponentially in $p$. 
This provides a setting to study localization in quantum Hall systems {\it without} critical states that are normally present due to the topological character.
Potentially, this could allow the electrons to exhibit many-body localization\cite{krishna_2017}.

\section{Numerical study of the plateau transition}
\label{sec:numerics}

\begin{figure}
\centering
\includegraphics[width=0.9\columnwidth]{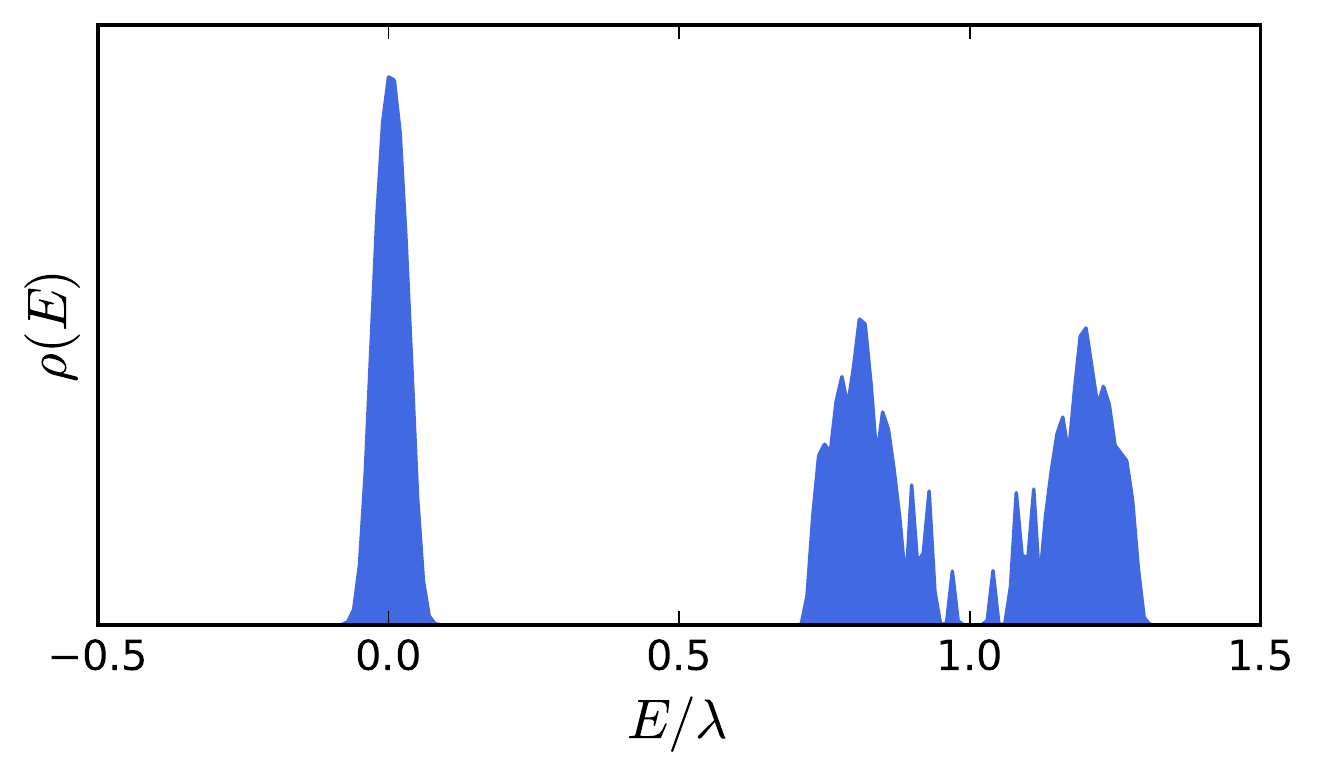}
\caption{
Density of states of a typical sample with 384 magnetic fluxes and 256 repulsive $\delta$-impurities arranged in a $16\times 16$ square lattice, in the lowest Landau level with the addition of weak disorder.
The singularity at $E=0$ (given by wavefunctions which vanish on all $\delta$-impurities) becomes a peak with finite width.
States in that peak have total Chern number $C=+1$ and thus display a quantum Hall transition as the Fermi energy is tuned across the peak.
The other states sit above a gap of order $O(\lambda)$ and have total Chern number $C=0$.
In this case, $p/q = 3/2$, there are two dispersive subbands at $E>0$ and one flat subband at $E=0$.
\label{fig:dos}}
\end{figure}

Since the electron states in $\LL(N_\phi, N_\delta)$ carry a total Hall current of $e^2/h$, one can study the integer quantum Hall transition in the disordered problem projected into that subspace.
More specifically, we consider the following Hamiltonian,
\begin{equation}
\mc H = \frac{1}{2m} \left( \mb p - \frac{e}{c} \mb A \right)^2 + V_n(\mb r) + \lambda V_\delta(\mb r)\;,
\label{eq:H_general}
\end{equation}
where $V_n(\mb r)$ is a disordered potential (e.g. Gaussian white noise \citep{huckestein_1995, huo_1992}), $V_\delta(\mb r)$ is a sum of repulsive $\delta$ function potentials, and $\lambda$ is a coefficient that determines the relative strength of the $\delta$-impurities compared to the disorder.
Once the Hamiltonian \eqref{eq:H_general} is projected into the lowest Landau level, the first term is reduced to the lowest Landau level energy $\frac{1}{2} \hbar \omega_c$, leaving
\begin{equation}
\mc H = \frac{1}{2} \hbar \omega_c + \mathcal P (V_n+ \lambda V_\delta) \mathcal P \;,
\label{eq:H_projected}
\end{equation}
$\mathcal P$ being the projector on the lowest Landau level.
Taking $\lambda$ to be much larger than the disorder strength, but still much smaller than the Landau level gap $\hbar \omega_c$ so as to avoid Landau level mixing, the density of states looks like the one depicted in Fig.~\ref{fig:dos}:
the $\delta$-function potential singularity at $E=\frac{1}{2} \hbar \omega_c$ gets broadened by the white noise to a narrow peak;
the rest of the states, with vanishing total Chern number, lie above a gap of order $\lambda$.

It is not obvious {\it a priori} how having some fraction of the wavefunction nodes pinned down at specified locations should affect 
various quantities, such as the Hall conductance $\sigma_{xy}(E_F)$ and the diagonal conductance $\sigma_{xx}(E_F)$ as a function of the Fermi energy $E_F$ (which we take to be within the density of states of the disordered potential). 
In particular, the localization length critical exponent characterizing the finite-size scaling behavior in the vicinity of the transition is a quantity of interest.

\subsection{Exact diagonalization of entire Landau level \label{sec:full}}

\begin{figure*}
\centering
\includegraphics[width = 0.9\textwidth]{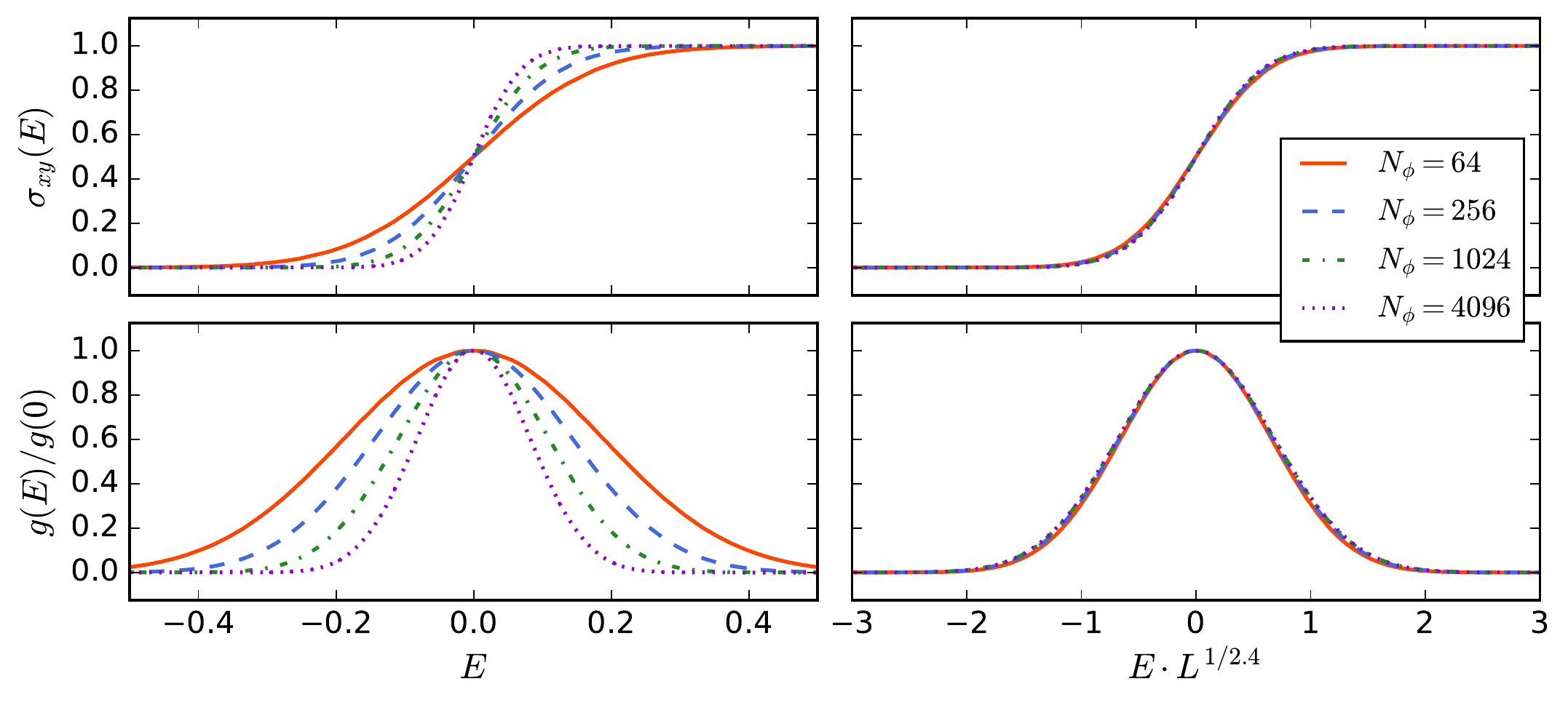}
\caption{Left: disorder-averaged conductances for a Landau level with $N_\phi$ flux quanta, $\LL(N_\phi,0)$. 
From top to bottom: Hall conductance $\sigma_{xy}$ and Thouless number $g$, defined in Eq.~\eqref{eq:thouless}, normalized to 1.
Right: the same quantities plotted as a function of $E\cdot L^{1/\nu}$.
$\sigma_{xy}$ and $g$ show scaling collapse for $\nu \approx 2.4$, consistent with the existing literature on this problem\cite{huckestein_1995, huo_1992}.
\label{fig:full}}
\end{figure*}

First, we consider the problem without $\delta$-impurities in the presence of white noise. 
We assume a square geometry with periodic boundary conditions.
We consider systems with $N_\phi = 4^n$, $n=3, 4, 5, 6$.
These values correspond to tori of linear size $L = 2^n \sqrt{h/eB}$. 
For each system size, we exactly diagonalize a large number of disorder realizations ($10^6$ for $N_\phi = 64$ and $256$, $10^5$ for $N_\phi = 1024$, and $10^4$ for $N_\phi = 4096$) to obtain the ensemble-averaged quantities.
Each disorder realization is represented by a Gaussian white noise potential $V_n(\mb r)$ of unit strength, projected into the lowest Landau level \citep{huckestein_1995}.
For each realization, we compute the Hall conductance via the Kubo formula
\begin{equation}
\sigma_{xy}(E) = \frac{e^2}{h} \nu(E) + \Delta \sigma_{xy}(E) \;,
\label{eq:filling}
\end{equation}
where $\nu(E)$ is the filling fraction when all states up to energy $E$ are occupied, and
\begin{align}
\frac{\Delta \sigma_{xy}(E) }{e^2/h}
& = - \frac{2}{N_\phi} \text{Im} \sum_{a<E_F<b} \frac{\bra{a} \partial_x V_n \ket{b} \bra{b} \partial_y V_n \ket{a} }{(E_a-E_b)^2} \;.
\label{eq:sigma}
\end{align}
The notation $a<E_F<b$ is shorthand for $a$ and $b$ such that $E_a < E_F < E_b$.
We also compute the Thouless number \citep{thouless_1974}:
\begin{equation}
g(E) = \frac{\langle |\delta E| \rangle_E}{\langle \Delta E \rangle_E}\;,
\label{eq:thouless}
\end{equation}
where $\Delta E$ is the level spacing and 
$\delta E$ is the change in energy of an eigenstate under a change of boundary conditions (periodic to anti-periodic) along one direction. 
$\langle \cdots \rangle_E$ denotes averaging over eigenstates around energy $E$.
The function $g(E)$ is related to the diagonal conductance, $\sigma_{xx}$, 
and is a measure of how localized a state is (low sensitivity to boundary condition changes is a signal of localization).

The disorder-averaged curves for $\sigma_{xy}$ and $g$, plotted in Fig.~\ref{fig:full}, 
show scaling collapse if $E$ is rescaled by $L^{-1/\nu}$, where the localization critical exponent is $\nu = 2.4 \pm 0.1$, consistent with known results\cite{huckestein_1995, huo_1992, thouless_1974} for these types of calculations.
More recent calculations\citep{slevin_2009, obuse_2012} with transfer matrix techniques on Chalker-Coddington network models suggest a larger exponent $\nu \approx 2.6$.
However, such large exponents are not seen in the continuum Landau level problem even with larger sizes\cite{bhatt_2018},
and recent work on disordered Chalker-Coddington network models\cite{gruzberg_2017} suggests a possible reason for the discrepancy.

\subsection{Projection onto kernel of $\delta$ function square lattice potential \label{sec:lattice}}

\begin{figure*}
\centering
\subfloat{
\includegraphics[width = 0.9\textwidth]{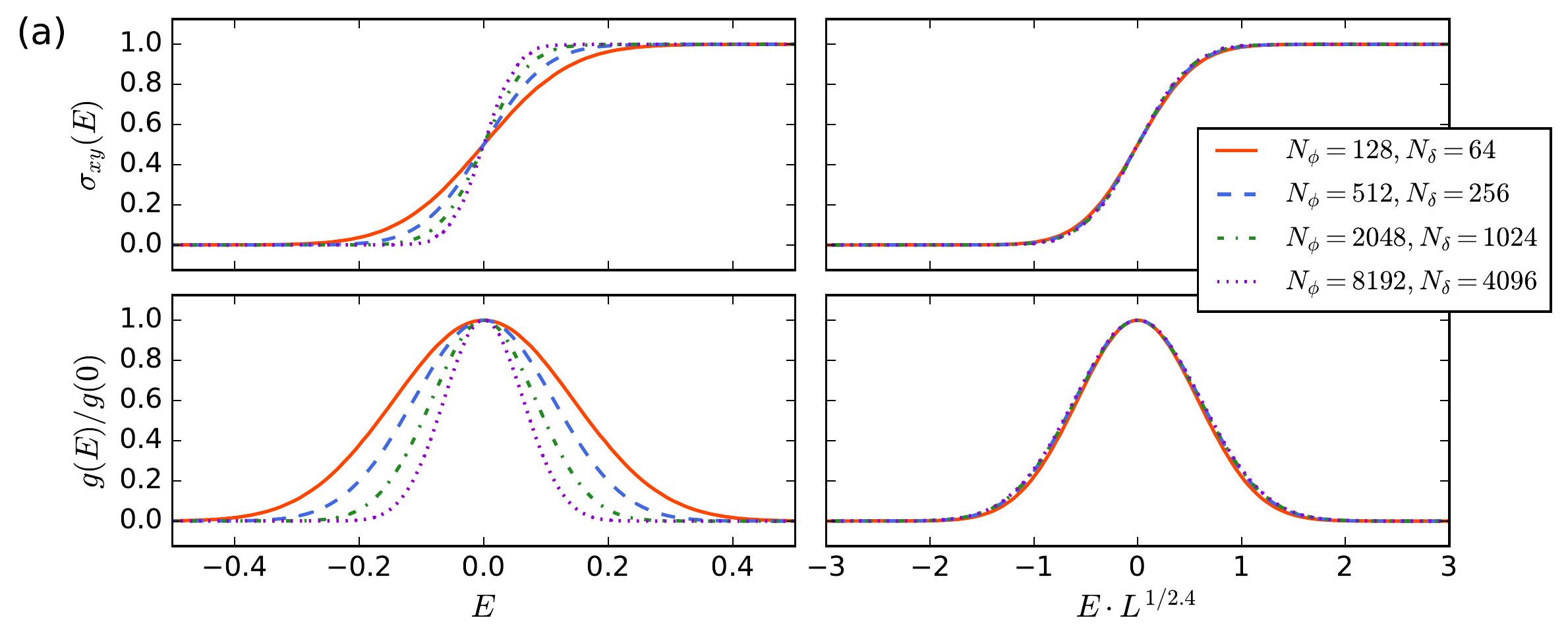}
} \\
\subfloat{
\includegraphics[width = 0.9\textwidth]{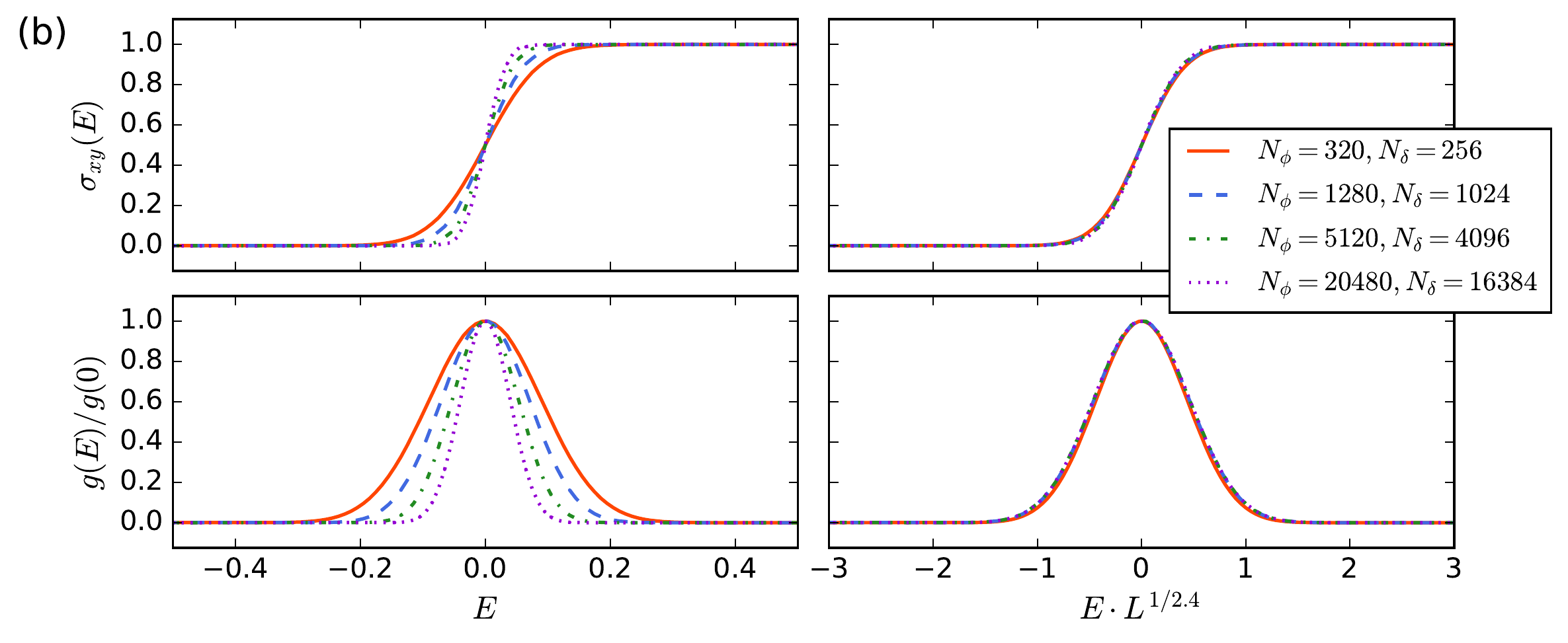}
}
\caption{(a)
Same plot as Fig.~\ref{fig:full}, but for a fraction of the lowest Landau level definied by $N_\phi$ flux quanta and $N_\delta = \frac{1}{2} N_\phi $ $\delta$-impurities, $\LL(N_\phi, N_\phi/2)$.
The choice of sizes is such that, for each curve, the truncated Hilbert space dimension ($N_\phi - N_\delta$) coincides with that of a curve in Fig.~\ref{fig:full}. 
(b)
Same plot, but with a larger truncation $N_\delta = \frac{4}{5} N_\phi $.
In both cases, the curves show scaling collapse with localization length critical exponent $\nu = 2.4\pm 0.1$.
\label{fig:square_scaling}}
\end{figure*}

We then add a square lattice of $\delta$ function potentials.
We consider systems with 2 magnetic fluxes per $\delta$ function, i.e. $p/q = 2$, and 5 magnetic fluxes per 4 $\delta$ functions, i.e. $p/q = 5/4$.
We fix $N_\phi - N_\delta = 4^n$, so that the dimension of $\LL(N_\phi, N_\delta)$ is $N_{\phi}' \equiv N_\phi - N_\delta = 4^n$, allowing for a straightforward comparison with the system sizes studied in the previous case, without $\delta$-impurities.
Lattice symmetry allows for an efficient diagonalization of this potential (see Appendix \ref{app:lattice}), and the resulting energy bands include one flat, zero-energy band and a dispersive band lying above a gap.

For each size, we diagonalize $V_\delta$ and obtain the projector $P$ onto its kernel $\LL(N_\phi, N_\delta)$, then generate a large number of disorder samples ($10^6$ for $N_\delta \leq 256$, $10^5$ for $N_\delta = 1024$, $10^4$ for $N_\delta = 4096$ and $2\times 10^3$ for $N_\delta = 16384$).
For each sample, we project the white noise potential $V_n$ and obtain the effective Hamiltonian $\mc H_{\rm trunc} = PV_nP$.
We diagonalize $\mc H_{\rm trunc}$ and compute $\sigma_{xy}$ and $g$ as in the previous case.
While the computation of the Thouless number translates straightforwardly, in the Hall conductance one cannot simply replace $V_n$ by the projected $PV_nP$ in Eq.~\eqref{eq:sigma}. 
Contributions coming from virtual hopping of electrons to the high-energy band of $V_\delta$ must be taken into account as well. 
Indeed, absent any extra terms, the Hall conductance of the zero-energy band, based on Eq.~\eqref{eq:filling}, would take the a value inconsistent with the Chern number.
A careful perturbative analysis (see Appendix \ref{app:kubo}) shows that $V_\delta$ induces terms in the Kubo formula which stay finite even in the $\lambda \to \infty$ limit of strong impurities.

The main benefit of using a lattice configuration of $\delta$ functions is computational speed:
lattice symmetry ensures $P$ has a sparse (block-diagonal) form, so that the additional manipulations required to compute $\sigma_{xy}$ are substantially faster than they would be for a generic distribution of $\delta$ functions (for which $P$ is generally dense).

Numerical results are shown in Fig.~\ref{fig:square_scaling}.
The data again exhibits scaling collapse with a localization length critical exponent $\nu = 2.4 \pm 0.1$.
Moreover, the functional forms of the Hall and Thouless conductances look the same as in the Landau level problem without $\delta$-impurities, Fig.~\ref{fig:full}.
On the other hand, if for each pair $(N_\phi, N_\delta)$ we compare the data for $\LL(N_\phi, N_\delta)$ and $\LL(N_\phi - N_\delta, 0)$ from the two figures, we see that in the former the slope of $\sigma_{xy}(E)$ is steeper and the peak in $g(E)$ is narrower.
This suggests the possibility that the observed behavior of $\LL(N_\phi, N_\delta)$ might capture some information about the plateau transition in a {\it larger} Landau level Hilbert space $\LL(N_\phi^{\rm eff}, 0)$, with $N_\phi \geq N_\phi^{\rm eff} \geq N_\phi - N_\delta$.
Following this hypothesis, by matching the width of the scaling function features, we find $N_\phi^{\rm eff} \approx 1.7 N_\phi$ at $p/q = 2$ and $N_\phi^{\rm eff} \approx 6.1 N_\phi$ at $p/q = 5/4$. 
This is manifestly unphysical, as the size of the entire Landau level, before projection, is only $N_\phi$: no information about the plateau transition in a Landau level of larger size is present in our scheme.

\begin{figure*}
\includegraphics[width = 0.75\textwidth]{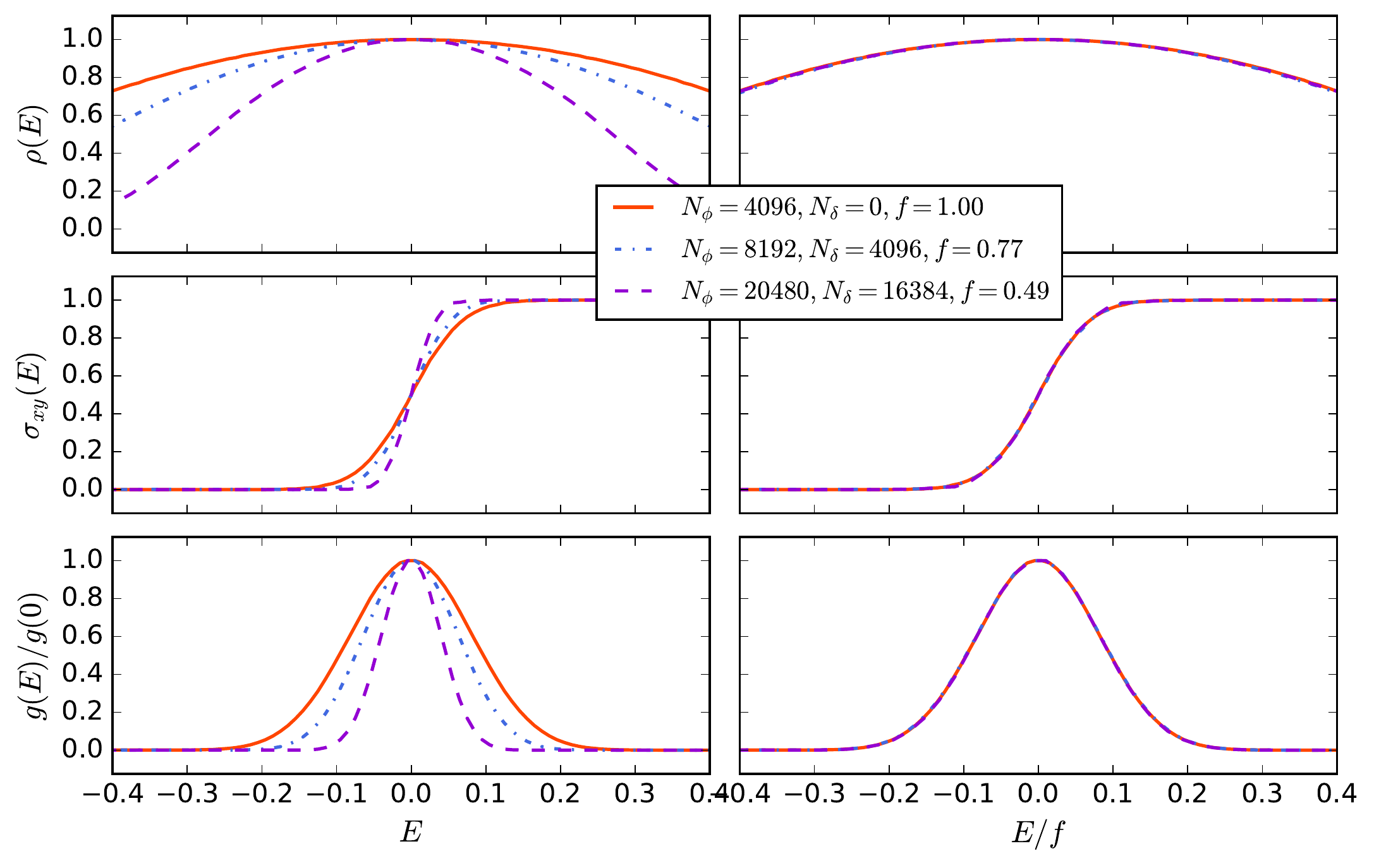}
\caption{
\label{fig:square_collapse}
Comparison between the plateau transition of a Landau level with 4096 flux quanta, $\LL(4096, 0)$, and that of two ``fractions'' of Landau levels $\LL(N_\phi, N_\delta)$, obtained from a square lattice of $\delta$-impurities, with fixed $N_\phi - N_\delta = 4096$.
Left: data for density of states $\rho$, Hall conductance $\sigma_{xy}$ and Thouless number $g$ as a function of energy $E$. 
Right: same data on a rescaled energy axis $E/f$ that matches the density of states in all three cases. 
The rescaling factor is $f = 1$ for $\LL(4096, 0)$, $f= 0.77$ for $\LL(8192, 4096)$, and $f = 0.49$ for $\LL(20480, 16384)$.
Upon rescaling the energy axis, $\sigma_{xy}$ and $g$ also coincide within numerical accuracy.
}
\end{figure*}

There is another explanation of the observed behavior, which arises more naturally by looking at the density of states.
In Fig.~\ref{fig:square_collapse} we compare the scaling functions and density of states for $\LL(N_\phi,N_\delta)$, at the two ratios we considered, and $\LL(N_\phi - N_\delta, 0)$.
For clarity, we only show the largest size we studied, $N_\phi - N_\delta = 4096$, as all other sizes yield analogous results.
We find that the presence of the $\delta$-impurities has a significant effect: 
it reduces the effective noise strength, and thus the width of the density of states.
Upon rescaling the energy axis to account for the change in effective disorder strength, we observe that the scaling functions $\sigma_{xy}$ and $g$ of $\LL(N_\phi, N_\delta)$ overlap with those of the whole Landau level without $\delta$-impurities, $\LL(N_\phi - N_\delta, 0)$.
This holds all the sizes we studied.

The renormalization of the noise strength can be understood heuristically as follows.
Localized orbitals are sensitive to the random potential averaged over a surface area $\sim \ell_B^2$,
\begin{equation}
\bar V_n (\ell_B) = \ell_B^{-2} \int_{r<\ell_B} d^2 \mathbf r V_n (\mathbf r)   \;.
\end{equation}
Assuming uncorrelated Gaussian white noise, such that $\langle V_n(\mathbf r_1) V_n(\mathbf r_2) \rangle = (aV_0)^2 \delta(\mathbf r_1 - \mathbf r_2)$, the variance of the locally averaged potential scales like $\ell_B^{-2}$:
\begin{align}
\langle \bar V_n (\ell_B)^2 \rangle 
& = \ell_B^{-4} \int_{r_1, r_2 < \ell_B} d^2 \mathbf r_1 d^2 \mathbf r_2 \langle V_n(\mathbf r_1) V_n(\mathbf r_2) \rangle \nonumber \\
& \sim \ell_B^{-4}  \int_{r_1<\ell_B} d^2 \mathbf r_1 (aV_0)^2  = (aV_0/\ell_B)^2 \;.
\label{eq:noise-average}
\end{align}
The pinned nodes of the wavefunctions act as magnetic flux tubes which let a fraction $q/p$ of the total magnetic field $B$ through the sample.
The remaining field $B(1-q/p)$ sets an effective magnetic length $\ell_B^* = (1-q/p)^{-1/2} \ell_B$.
Thus localized orbitals average the white noise over a larger area and, based on Eq.~\eqref{eq:noise-average}, experience a {\it reduced} disorder strength $V_{\rm eff} = V(1-q/p)^{1/2} $.
This simple derivation is in qualitative agreement with what we observed for a variety of $p/q$ ratios; for the cases discussed in this Section ($p/q=2$ and $p/q = 5/4$) it gives $1/\sqrt{2} \simeq 0.71$ and $1/\sqrt{5} \simeq 0.45$, reasonably close to the observed values of $0.77$ and $0.49$. 

Remarkably, up to the aforementioned rescaling of the disorder strength (or, equivalently, of the magnetic length), the plateau transition in a fraction of the lowest Landau level looks exactly the same as in the full lowest Landau level of corresponding dimension.
This result suggests that, besides the localization length critical exponent, 
{\it quantitative} details of the plateau transition
depend solely on the dimensionality of the Hilbert space of the flat, $C=1$ band (in this case a fraction of the lowest Landau level). 
They do not seem to depend on other details of the states that make up such band.

\subsection{Different spatial distributions of $\delta$ functions \label{sec:check}}

In order to check that our result is general and not a special property of the square lattice distribution previously considered, we performed checks with different geometric distributions of $\delta$ functions.

\begin{figure}
\centering
\includegraphics[width = \columnwidth]{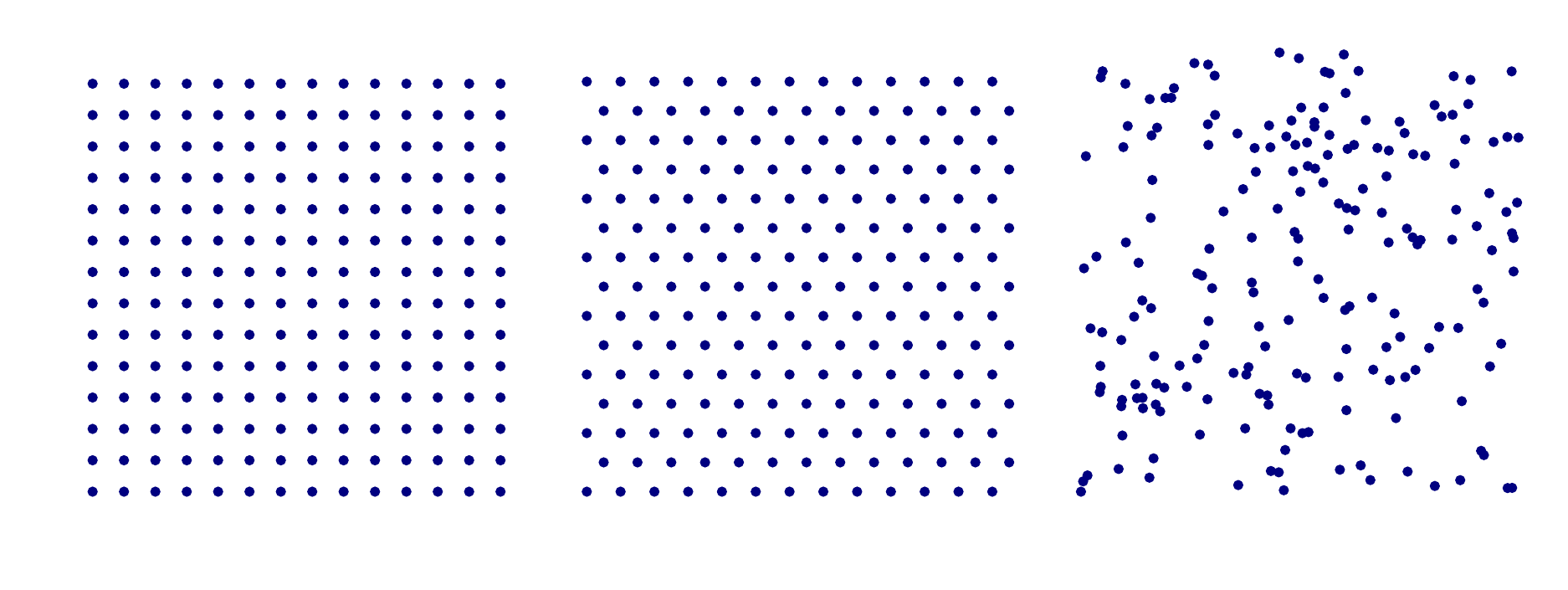}
\caption{\label{fig:deltas}
The three distributions of $\delta$-impurities we consider in Sec.~\ref{sec:check}.
Left: a $14\times 14$ square lattice. 
Center: a $13\times 15$ triangular lattice.
Right: 195 randomly distributed $\delta$-impurities. Each position is sampled out of a uniform distribution on the torus and discarded if within $0.5 \ell_B$ of any other $\delta$-impurity.
}
\end{figure}

First we considered a triangular lattice.
Given the irrational aspect ratio of the triangular lattice unit cell, it is impossible to accommodate an integer number of $\delta$ functions on a square torus in a perfectly triangular lattice.
Thus either the torus or the lattice configuration must be made slightly anisotropic.
We use an anisotropic triangular lattice with unit cell aspect ratio of $13/15$, which approximates $\sqrt{3}/2$ to one part in $10^3$.
We thus fit $13\times15 = 195$ $\delta$-impurities on a square torus in a nearly triangular lattice and study the plateau transition in the subspace $\LL(390, 195)$ by diagonalizing $10^6$ disorder realizations.

Then we consider a random distribution of $\delta$ functions, again in a system with $N_\phi = 390$ and $N_\delta = 195$. 
We sample the positions from a uniform distribution on the square torus, but require any two $\delta$-impurities to be at leas $0.5\ell_B$ away from each other (for comparison, the spacing in the lattices we considered is $\sim 3\ell_B$) to ensure that the projection on $\LL(N_\phi, N_\delta)$ is well-behaved\footnote{If two $\delta$-impurities get very close to each other, $V_\delta$ can have very small, non-zero eigenvalues arising from wavefunctions that have a node on one $\delta$-impurity but not on the other.}.
We generate $10^2$ such configurations and, for each one, we diagonalize $10^4$ disorder realizations.

\begin{figure}
\centering
\includegraphics[width = \columnwidth]{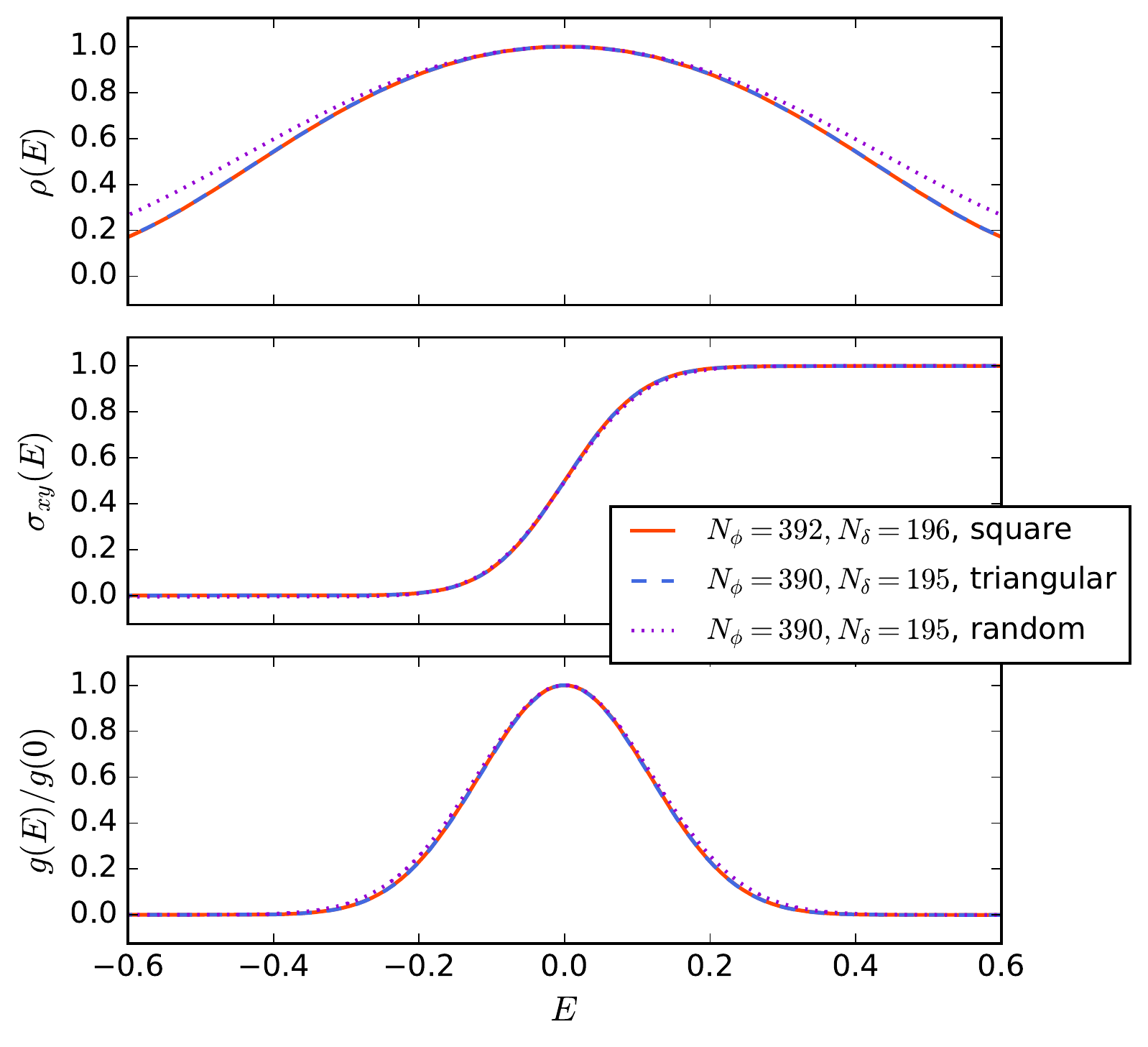}
\caption{ \label{fig:check}
Comparison between plateau transitions in the $E = 0$ energy subspace of the three $\delta$-impurity potentials represented in Fig.~\ref{fig:deltas}: a square lattice, a triangular lattice, and a random distribution.
$10^6$ disorder realizations are averaged for each case (for the random case, $10^2$ distributions of $\delta$-impurities are considered and for each of them $10^4$ realizations of white-noise disorder are averaged).}
\end{figure}

These two cases can be compared to the transition in the \textit{square} lattice problem $\LL(392, 196)$, where the $\delta$-impurities are arranged on a $14\times 14$ square lattice\footnote{The difference in electron number between the square and triangular cases is expected, based on the scaling hypothesis, to cause a small discrepancy of order $1/2\nu N_\phi \sim 10^{-3}$ in the collapse of the scaling functions between the two cases.},
which we know is equivalent to $\LL(195, 0)$ from the analysis presented in the previous section.
For this case, too, we average $10^6$ disorder realizations.
The results are shown in Fig.~\ref{fig:check}.
For the two lattices, the density of states and scaling functions overlap within uncertainty at all energies.
For the random distribution, the density of states has larger tails and the scaling functions also show small deviations away from $E = 0$.
These tails arise from the spatial non-uniformity of the random distribution, as can be seen in Fig.~\ref{fig:deltas}:
states localized in regions without $\delta$-impurities, or with an abnormally low concentration of them, experience a stronger effective disorder.
Nonetheless, in the bulk of the critical regime, the forms of $\rho$, $g$ and $\sigma_{xy}$ coincide for all distributions.
This quantitative match implies universality that goes beyond the critical exponent:
the diagonal and Hall conductivities match quantitatively, independently of the geometric distribution of $\delta$ functions.
This further suggests that the plateau transition is the same for any $C=1$ flat band of a given dimensionality, be it a whole Landau level or a fraction of a larger one.

\section{Conclusions and discussion \label{sec:conclusion}}

We have presented a method to isolate a ``fraction'' of the lowest Landau level (LLL) by using $\delta$-function potentials. 
This subspace shares many features with the original LLL, including its total topological Chern character and its vanishing bandwidth. 
It therefore undergoes a quantum Hall plateau transition, but its lower dimensionality makes it easier to study numerically. 
Physically, each $\delta$-impurity effectively binds, or localizes, one electronic state, so that $N_\delta$ impurities in an $N_\phi$-dimensional LLL give rise to a subspace $\LL(N_\phi, N_\delta)$ of dimension $N_\phi' \equiv N_\phi-N_\delta$ consisting of electron states that avoid all the $\delta$-impurities.

We studied the integer quantum Hall plateau transition in this subspace by exact diagonalization of large numbers of disorder realizations.
Our numerical results indicate that the transition in the ``fraction'' of the LLL $\LL(N_\phi,N_\delta)$ \textit{quantitatively} matches the one occurring in $\LL(N_\phi-N_\delta, 0)$, i.e. the whole LLL of a system with $(N_\phi-N_\delta)$ magnetic flux quanta and no $\delta$-impurities.
The only effects of the $\delta$-impurities are 
(i) an effective reduction of the magnetic field through the system, or equivalently an effective increase in the magnetic length; and, as a consequence of that,
(ii) an effective reduction in the strength of the disorder.

From this we conclude that the plateau transition, 
and in particular the localization length critical exponent, 
is the same for all flat $C=1$ bands with a given number of states.
Our results thus suggest that a computational speedup for finite-size scaling studies of the plateau transition cannot be achieved by retaining only a fraction of the LLL.
It remains to be seen whether or not this conclusion changes in the presence of interactions:
the many-body problem in the truncated Landau level $\LL(N_\phi, N_\delta)$ may reveal some information about the problem in the {\it larger} Landau level $\LL(N_\phi, 0)$,
or it may map exactly to the problem in the {\it smaller} Landau level $\LL(N_\phi - N_\delta, 0)$.
Both options have interesting consequences.

In the first case, a truncation of the single-particle Hilbert space would lead to an {\it exponential} computational speedup for the many-body problem, potentially allowing exact diagonalization studies of unprecedented system sizes.
The latter case, on the other hand, would potentially enable interesting experimental applications.
Attractive point-like impurities could be used to artificially reduce the density of carriers in quantum Hall systems: 
by binding a fraction $0<f<1$ of the carriers into inert, topologically trivial states at large negative energies, one could obtain effective filling fractions $\nu^*$ of the flat $C=1$ band with a larger overall LLL filling fraction of $\nu = f + \nu^* (1-f)$. 
This could allow the observation of fractional quantum Hall states in systems where that would otherwise require exceedingly high magnetic fields.
In this respect, recent progress towards the realization of artificial super-lattices in various two-dimensional electron systems\cite{dean_2017, manfra_2017} appears promising.

\acknowledgments

We thank Michael Zaletel and Akshay Krishna for useful conversations.
This work was supported by DOE grant DE-SC0002140.

\appendix

\section{Direct computation of the Chern number}
\label{app:integral}

In this Appendix we directly compute the Chern number of the flat band of electrons at $E = \frac{1}{2} \hbar \omega_c$ in the presence of $N_\phi$ quanta of magnetic flux through the system and $N_\delta$ $\delta$-function impurities, via the boundary integral
\begin{equation}
C = \frac{1}{2\pi i} \oint d\theta_i \langle \Psi_0 (\vec \theta) | \frac{\partial}{\partial \theta_i} | \Psi_0(\vec \theta) \rangle \;.
\end{equation}
The key observation is that the $\vec\theta$ dependence is only in the center-of-mass part of the wavefunction, $F_{cm}(Z; \vec \theta)$.
This allows us to rewrite the integral as
\begin{align}
C
& =  \frac{1}{2\pi i} \oint d\theta_i \int d^{N_{\phi}'} x\, d^{N_{\phi}'} y \,
| \Psi_0(\{z_i\}; \vec \theta) |^2  \nonumber \\
& \qquad \times \frac{\partial}{\partial \theta_i}  \log F_{cm} (Z; \vec \theta)
\end{align}
It is useful to split the integral into the four sides of the rectangle.
Since a $2\pi$ change in the boundary angles can cause at most a phase shift, the weight $| \Psi_0(\{z_i\}; \vec \theta) |^2$ is the same for corresponding points on opposite sides.
So pairs of opposite sides can be grouped as follows:
\begin{align}
C
& =  \frac{1}{2\pi i}\int_0^{2\pi} d\theta_x \int d^{2N_{\phi}'} z_i \, \nonumber \\
& \left(  | \Psi_0(\{z_i\}; \theta_x,0) |^2 
\frac{\partial}{\partial \theta_i}  
\log \frac{ F_{cm} (Z; \theta_x,0) }{ F_{cm} (Z; \theta_x,2\pi) } \right. \nonumber \\
& \left. + | \Psi_0(\{z_i\}; 2\pi,\theta_y) |^2 
\frac{\partial}{\partial \theta_i}  
\log \frac{ F_{cm} (Z; 2\pi, \theta_y ) }{ F_{cm} (Z; 0, \theta_y) } \right) \;.
\end{align}
The logarithm of the ratio of $F_{cm}$ functions yields two terms:
the first is $\pm \frac{2\pi}{L_x} Z$, which is independent of $\theta$ and thus vanishes when we take the derivative;
the second only includes the $\vartheta_1$ functions.
Exploiting the quasiperiodicity of $\vartheta_1$ we finally get
\begin{align}
C
& = \frac{1}{2\pi i} \int_0^{2\pi} d\theta\, 
\int d^{N_{\phi}'}xd^{N_{\phi}'} y\, 
| \Psi_0(\{z_i\}; 0,\theta) |^2 \nonumber \\ 
& \qquad \times \frac{\partial}{\partial \theta} 
\left(\pi \left(i+ \frac{L_y}{L_x} \right) -2\pi i \frac{Z}{L_x} + i \theta \right) \nonumber \\
& =\int_0^{2\pi} \frac{d\theta}{2\pi} \int d^{N_{\phi}'}xd^{N_{\phi}'} y\, 
| \Psi_0(\{z_i\}; 0,\theta) |^2 = +1\;.
\end{align}

\section{Diagonalization of the $\delta$ function lattice potential \label{app:lattice}}

In this Appendix we discuss the diagonalization of the $\delta$-function potential when the impurities are arranged on a lattice.
We consider the potential introduced in Eq.~\eqref{eq:V_of_r} and its block-diagonal structure with respect to quasi-momentum eigenstates, Eq.~\eqref{eq:vk_block}.

Let us pick a value of the quasi-momentum $\mb k$ and compute the corresponding block of the potential matrix,
\begin{align}
V_{\delta}^{ab} (\mb k) 
= \langle \psi_{\mb k,a} | V | \psi_{\mb k,b} \rangle 
& = \lambda \sum_{n_1 = 1}^{N_1} \sum_{n_2 = 1}^{N_2} 
\psi_{\mb k,a}^*(n_1 \mb a_1 + n_2 \mb a_2) \nonumber \\
& \quad \times \psi_{\mb k,b}(n_1 \mb a_1 + n_2 \mb a_2)\;.
\end{align}
We can replace the $\psi$ functions with the pseudo-Bloch $u$ functions, since the plane-wave phase factors cancel out.
Furthermore, by using Eq.~\eqref{eq:u}, 
we obtain that all terms related by a magnetic-unit-cell translations are identical, 
and thus the sum reduces to
\begin{align}
V_{\delta}^{ab} (\mb k) 
& = \langle \psi_{\mb k,a} | V_\delta | \psi_{\mb k,b} \rangle  \nonumber \\ 
& = \lambda \frac{N_\delta}{q} \sum_{n=1}^q 
u_{\mb k,a}^*(n \mb a_1) u_{\mb k,b}(n \mb a_1)\;.
\label{eq:V_MUC}
\end{align}
By defining the $q \times p$ matrix $v_{n, a}(\mb k) \equiv u_{\mb k, a}(n\mb a_1)$, 
we get the form $V(\mb k) \propto v(\mb k )^\dagger v(\mb k)$ discussed in the main text.

As an example, for a square lattice, one has
\begin{align}
u_{\mb k,a}(n \mb a_1) 
& = \frac{1}{\sqrt{\mc N}} \vartheta_3 \left( \frac{a}{p} - \frac{n}{q} - i l \frac{k_x+ik_y}{2\pi p} \bigg| \frac{i}{pq} \right), \label{eq:u_theta3}
\end{align}
where $\mc N$ is a normalization factor independent of $a$ (but  generally dependent on $\mb k$), $l = |\mb a_1|$ is the lattice spacing, and $\vartheta_3$ is the third elliptic Theta function, 
\begin{equation}
\vartheta_3(z|\tau) = \sum_{l \in \mathbb Z} e^{i \pi \tau l^2} e^{-2\pi i l z}\;.
\end{equation} 
Thus at any given quasi-momentum the Hamiltonian block in Eq.~\eqref{eq:V_MUC} is 
\begin{align}
V_{\delta}^{ab} (\mb k) 
= & 
\frac{1}{\mc N} \sum_{n = 1}^q 
 \ {\vartheta_3 \left( \frac{a}{p} - \frac{n}{q} - i l \frac{k_x+i k_y}{2\pi p} \bigg | \frac{i}{pq} \right) }^* \nonumber \\
&  \ \times \vartheta_3 \left( \frac{b}{p} - \frac{n}{q} - i l \frac{k_x+i k_y}{2\pi p} \bigg | \frac{i}{pq} \right) \;.
\label{eq:vmn_k}
\end{align}

The resulting bands are found by diagonalizing $V_\delta(\mb k)$, which is a $p \times p$ matrix of rank $q$.
As such, it has $q$ non-zero eigenvalues and a kernel of dimension $p-q$.
Practically, the projector $P$ on the kernel of $V_\delta$,
which we use in Section~\ref{sec:lattice},
is obtained by diagonalizing $V_\delta(\mb k)$ in the basis of quasi-Bloch wavefunctions as discussed earlier, and then transforming to the basis of usual Landau orbitals on a torus, 
$$
\phi_n(x,y) \propto \sum_{p \in \mathbb Z} e^{2\pi y (n+N_\phi p)/L} e^{-\frac{1}{2} (x+\frac{n}{N_\phi} L + pL)^2} \;.
$$
The whole process takes computational time $O(N_\phi p^2)$,
as opposed to the numerical diagonalization of a generic $\delta$-function potential without lattice symmetry which takes $O(N_\phi^3)$.
Furthermore, the resulting projector is sparse due to the quasi-momentum quantum number, which speeds up the calculation of $PV_nP$.

\section{Kubo formula for $\sigma_{xy}$ of states in the zero-energy subspace of the $\delta$-impurity potential \label{app:kubo}}

In this Appendix we discuss the Kubo formula for the Hall conductance of states in the kernel of a $\delta$-function potential, in the limit of very strong $\delta$-functions.

The Hamiltonian for the full system is 
\begin{equation}
\mc H = \frac{1}{2} \hbar \omega_c + V_n + \lambda V_{\delta}\;,
\end{equation}
where $V_n$ is a Gaussian white noise potential of strength $V_0$ (we set $\ell_B$ as the unit of length), 
$$ \langle V_n(\mb r) V_n(\mb r') \rangle = V_0^2 \delta(\mb r - \mb r')\;, $$
and
$$V_\delta(\mb r) = V_0 \sum_{j=1}^{N_\delta} \delta(\mb r- \mb R_j)\;.$$
The set of $\delta$-potentials locations $\{\mb R_j\}$ is arbitrary.
We assume $\lambda \gg 1$ and work in perturbation theory in the small parameter $\eta \equiv 1/\lambda \ll 1$.
For convenience, we drop the constant $\frac{1}{2} \hbar \omega_c$ (which does not affect the Hall conductance), rescale $\mc H$ by $\lambda$ and define $V \equiv V_\delta + \eta V_n$.

The Kubo formula for the exact eigenstate \ket{\psi_a} of $V$ reads
\begin{equation}
\sigma_{xy}(a) = \frac{e^2}{N_\phi h} + \Delta \sigma_{xy}(a)\;,
\end{equation}
with
\begin{equation}
\frac{\Delta \sigma_{xy}(a)}{e^2/h} = -2\text{Im} \sum_{b \neq a} \frac{ \bra{\psi_a} \partial_x V \ket{\psi_b} \bra{\psi_b} \partial_yV \ket{\psi_a} }{(E_a-E_b)^2} \;.
\label{eq:kubo}
\end{equation}
We are interested in states $a$ in the kernel of $V_\delta$, so in the perturbative expansion
$$\ket{\psi_a} = \ket{\psi_a^{(0)}} + \eta \ket{\psi_a^{(1)}} + \dots$$ 
one has $V_\delta \ket{\psi_a^{(0)}} = 0$.
Sorting the states based on increasing energy, the kernel of $V_\delta$ corresponds to $a\leq N_\phi-N_\delta$.

We expand the Kubo formula using perturbation theory in $\eta \ll 1$ and retain all contributions of order 1.
There are two types of terms in the sum in Eq.~\eqref{eq:kubo}: those with $b> N_\phi - N_\delta$, and those with $b\leq N_\phi-N_\delta$.
In the former case case, the denominator is $O(1)$, so only terms of $O(1)$ in the numerator matter.
The only such term is $\bra{\psi_a^{(0)}} \partial_x V_\delta \ket{\psi_b^{(0)}} \bra{\psi_b^{(0)}} \partial_y V_\delta \ket{\psi_a^{(0)}} $.
So the sum over $b>N_\phi-N_\delta$, in the limit $\eta \to 0$, is
\begin{align}
& \bra{\psi_a^{(0)}} \partial_x V_\delta 
\left( \sum_{b>N_\phi - N_\delta}  \frac{ \ket{\psi_b^{(0)}} \bra{\psi_b^{(0)}} }{E_b^2}  \right) 
\partial_y V_\delta \ket{\psi_a^{(0)}}  \nonumber \\
& = \bra{\psi_a^{(0)}} \partial_x V_\delta Q V_\delta^{-2} Q \partial_y V_\delta \ket{\psi_a^{(0)}} \;,
\end{align}
where $Q$ is the orthogonal complement to $P$, the projector on the kernel of $V_\delta$.

In the latter case, i.e. if $b \leq N_\phi - N_\delta$, 
$E_b$ is $O(\eta)$ like $E_a$, so that the energy denominator is $O(\eta^2)$.
For the limit $\eta \to 0$ to be finite, all terms $O(1)$ or $O(\eta)$ must vanish in the numerator.
This is indeed the case, since 
\begin{equation}
\bra{\psi_a^{(0)}} \partial_i V_\delta \ket{\psi_b^{(0)}} 
= \sum_{j=1}^{N_\delta} \left. \partial_i ( \psi_a^{(0)*} \psi_b^{(0)} ) \right|_{\mb r=\mb R_j} = 0
\end{equation}
as the product $\psi_a^{(0)*}(\mb r) \psi_b^{(0)} (\mb r) $ has a double zero at all impurity locations. 
This leaves, as the next leading terms, products of pairs of the following terms, all $O(\eta)$:
\begin{align}
\bra{\psi_a^{(1)}} \partial_x V_\delta \ket{\psi_b^{(0)}} 
& = - \bra{\psi_a^{(0)}} V_n Q V_\delta^{-1} Q \partial_x V_\delta \ket{\psi_b^{(0)}} \;, \\
\bra{\psi_a^{(0)}} \partial_x V_\delta \ket{\psi_b^{(1)}} 
& = - \bra{\psi_a^{(0)}} \partial_x V_\delta Q V_\delta^{-1} Q V_n \ket{\psi_b^{(0)}} \;, \\
\bra{\psi_a^{(0)}} \partial_x V_n \ket{\psi_b^{(0)}}
& \;.
\end{align}
This is simplified by introducing the operators 
\begin{equation}
\Delta_i \equiv Q V_\delta^{-1} Q \partial_i V_\delta \;.
\end{equation}

Putting the two contributions together, 
the overall result to $O(1)$ in the limit $\eta \to 0$ from the sum over all $1 \leq b \leq N_\phi$ is 
\begin{widetext}
\begin{equation}
\frac{\Delta \sigma_{xy}(a)}{e^2/h} = 
-\frac{2}{N_\phi}\text{Im} \bra{\psi_a^{(0)}} \left(
 \Delta_x^\dagger \Delta_y +
\sum_{\substack{b<N_\phi-N_\delta \\ b\neq a}} 
\frac{(\partial_x V_n -V_n\Delta_x -\Delta_x^\dagger V_n ) \ket{\psi_b^{(0)}} 
\bra{\psi_b^{(0)}}   (\partial_y V_n -V_n\Delta_y -\Delta_y^\dagger V_n )  }{(E_a^{(0)} - E_b^{(0)} )^2}  \right) \ket{\psi_a^{(0)}}  \;,
\label{eq:sigma_correction1}
\end{equation}
\end{widetext}
where all the disorder-dependent data comes from the {\it projected} problem:
the $\ket{\psi_a^{(0)}}$ and $E_a^{(0)}$ are respectively eigenvectors and eigenvalues of the projected potential $PV_nP$. 
The $\Delta_i$ operators do not depend on the disorder realization $V_n$, and thus must be computed only once.
This guarantees that the computation of $\sigma_{xy}$ in $\LL(N_\phi, N_\delta)$ is almost as efficient as that of $\sigma_{xy}$ in $\LL(N_\phi - N_\delta, 0)$, with only a small overhead.

\bibliography{iqht_fractionLL}

\end{document}